\documentclass{nature}

\usepackage{amsmath,amssymb,amsfonts,graphicx,bm,color}
\usepackage{mathrsfs}
\usepackage[utf8]{inputenc}
\usepackage[T1]{fontenc}
\usepackage{lmodern}

\title{Graphene-based autonomous pyroelectric system for near-field energy conversion}

\author{I. Latella$^1$ and  P. Ben-Abdallah$^{2}$}

\makeatletter
\let\saved@includegraphics\includegraphics
\AtBeginDocument{\let\includegraphics\saved@includegraphics}
\renewenvironment*{figure}{\@float{figure}}{\end@float}
\makeatother

\begin{document}

\maketitle

\begin{affiliations}
 \item Departament de F\'{i}sica de la Mat\`{e}ria Condensada, Universitat de Barcelona, Mart\'{i} i Franqu\`{e}s 1, 08028
Barcelona, Spain.
 \item Laboratoire Charles Fabry,UMR 8501, Institut d'Optique, CNRS, Universit\'{e} Paris-Sud 11,
2, Avenue Augustin Fresnel, 91127 Palaiseau Cedex, France.
\end{affiliations}


\begin{abstract}
In the close vicinity of a hot solid, at distances smaller than the thermal wavelength, a strong electromagnetic energy density exists\cite{Eckardt} because of the presence of evanescent field. Here we explore the possibility to harvest this energy using graphene-based pyroelectric conversion devices made with an active layer encapsulated between two graphene field-effect transistors\cite{Novoselov} (GFETs) deposited on the source and on the cold sink. By tuning the bias voltage applied to the gates of these transistors, the thermal state and the spontaneous polarization of the active layer can be controlled at kHz frequencies. We demonstrate that the power density generated by these conversion systems can reach $1300\:\mathrm{W.\,m}^{-2}$ using pyroelectric Ericsson cycles, a value which surpasses the current production capacity of near-field thermophotovoltaic conversion devices~\cite{DiMatteo, Narayanaswamy,Fiorino,Bhatt} by more than five orders of magnitude with low grade heat sources ($T<500\,\mathrm{K}$) and small temperature differences ($\Delta T\sim 100\,K$). 
\end{abstract}

Over 60\,\% of energy used in industry is lost as low grade waste heat, which constitutes an important source of renewable energy available. Unfortunately, few thermodynamic cycles can operate at low temperature streams effectively\cite{DiSalvo}. Thermophotovoltaic (TPV) generation of electricity\cite{Lenert} from heat flux radiated  by hot sources was expected to play a major role in harvesting this waste heat. However the energy flux exchanged between the primary source and the cell in these devices is intrinsically limited by the Stefan-Boltzmann's law (i.e. the heat flux exchanged between two blackbodies), setting so a relatively low upper bound\cite{Coutts} for the generated power. In the 70's, an important step forward has been taken to improve the performances of this technology. Indeed, when the source and the cell are separated by a subwavelength distance, non-propagating photons can be transferred to the cell by tunneling effect\cite{Polder,Ottens,Pramod1,St-Gelais}, carrying a heat flux that can largely surpass the limit set by the blackbody theory. This energy transfer in the near-field regime paved the way for a novel technology, the so-called near-field thermophotovoltaic (NTPV) energy conversion. Although theoretically this technology can lead to a generated power of about $1\,$W.\,cm$^{-2}$ with heat sources at $T>500\,$K and separation distances with the cell of few hundred nanometers (10\,W.\,cm$^{-2}$ being today the energy demand of a household in US), several hurdles still limit today its development and massive deployment. One of the main limitations is the mismatch between the emission frequency of evanescent modes supported by the source and the gap frequency of the junction, the frequency below which the photon energy is dissipated as heat into the cell,  reducing dramatically its performance. So, despite its theoretical potential, only generated powers of few $\mu$W.\,cm$^{-2}$ have been reported\cite{Fiorino,Bhatt} so far with such devices.

In 2010 Fang et al.\cite{Fang} proposed an alternative way to harvest the near-field thermal energy by using a pyroelectric converter. In this technology, an active layer made with a pyroelectric material undergoes a temporal variation of its temperature thanks to a periodic modulation of its separation distance with the hot source and the cold sink. With a distance of $100\,$nm, an operating frequency of few Hz and an electric power of $6.5\,$mW$.$cm$^{-2}$ have been predicted when a hot source is at $T_1=383\,$K and a cold sink at $T_3=283\,$K. Improving the performances of these converters by increasing their operating frequency up to kHz without reducing the amplitude of the temperature modulation of the active layer could in principle make them competitive with the NTPV technology and could even surpass it. 
However, this remains today a challenging problem, since it requires working with thin active layers at smaller separation distances from the source and the sink, distances for which the Casimir force induced by the vacuum fluctuations limits the possibility of moving the active layer with a reasonable energy consumption (this force per unit area is of the order of $10\,$N$.$m$^{-2}$ for a separation of $100\,$nm and it increases to $10^5\,$N$.$m$^{-2}$ at $10\,$nm).
Here we address these challenges by introducing a static (non-mechanical) pyroelectric converter based on graphene (Fig.\,1a), whose properties can be externally controlled with an applied voltage in order to modulate at kHz frequencies the near-field interactions between the pyroelectric membrane and both the source and the sink. By dynamically controlling the charge density of the graphene sheets and exploiting the tunability of surface wave coupling between the different elements of the converter, we demonstrate that these pyroelectric devices can generate an electric power larger than a hundred mW$.$cm$^{-2}$ with low grade heat sources. 
Moreover, on the contrary to solid-state pyroelectric converters operating at kHz frequencies\cite{Bhatia,Pandya}, we demonstrate that our graphene-based pyroelectric system is a self-powered or autonomous conversion device in which the power required to modulate the temperature is much smaller than the delivered power, opening so a new avenue for high-frequency pyroelectric energy harvesting from stationary thermal sources.


The proposed device consists in an active membrane made with a pyroelectric layer of thickness $\delta_p$ which is covered on both sides by a gold (Au) layer, acting as electrode, and a superficial silica (SiO$_2$) layer which supports a surface wave in the infrared. The electrodes are taken sufficiently thick (here $200\,$nm) in order to screen the incoming electromagnetic field in the infrared, while the thickness of SiO$_2$ layers is chosen small enough (here $50\,$nm) to limit the thermal inertia of the active membrane. 
As sketched in Fig.\,1a, this membrane (body 2) at temperature $T_2$ is encapsulated without contact between a hot source (body 1) and a cold sink (body 3) at temperatures $T_1$ and $T_3$, respectively. These two thermal reservoirs consist in a multilayer structure made with a $n$-doped silicon (Si) substrate surmounted by a SiO$_2$ layer of thickness $\delta_g=5\,$nm which is itself covered by a graphene sheet, the whole constituting a GFET. By applying an external bias voltage $V_{gi}$ on the gate of each of these GFETs  operating in cut-off mode (i.e. no current flows  from the GFET source to the drain), the superficial carrier density\cite{Novoselov} $n_{gi}=C_gV_{gi}/e$ on the graphene sheet and therefore its chemical potential  $\mu_{gi}=\hbar v_F\sqrt{\pi n_{gi}}$ can be actively controlled (here $e$ is the electron charge, $\hbar$ is the reduced Planck constant, $v_F=10^6\,$m$/$s is the Fermi velocity and $C_g=\varepsilon_g/\delta_g$ is the capacitance per unit surface of the GFET, $\varepsilon_g$ being the permittivity of the dielectric layer). It follows that the radiative coupling between the active membrane and the two thermal reservoirs can also be dynamically tuned with the modulation of these bias voltages. 

In contrast to the Fang et al. converter\cite{Fang}, in our three-terminal device the separation distance between the active membrane and the two reservoirs is kept fixed and equal to $d=20\,$nm while the bias voltages $V_{g1}$ and $V_{g2}$ applied on the GFETs undergo periodic rectangular modulations at frequency $f$ in phase opposition. 
According to the theory of radiative heat transfer in many-body systems\cite{pbaPRL2011,Latella}, the net power per unit surface received by the active layer reads
\begin{equation}
\mathcal{P}_{r}(V_{gi};T_2,t)= \int_0^\infty \frac{d\omega}{2\pi}\, \varphi_{12}(\omega) -\int_0^\infty \frac{d\omega}{2\pi}\, \varphi_{23}(\omega).
\label{Eq:Prad}
\end{equation}
Here $\varphi_{mn}(\omega)= \theta_{mn}(\omega)\sum_{l = \{\rm TE,TM\}} \int \frac{d^2\mathbf{k}}{(2\pi)^2} \mathcal{T}_l^{mn}(\omega,k)$ represents the spectral flux at frequency $\omega$, weighted by the transmission coefficient  $\mathcal{T}_l^{mn}(\omega,k)$ which describes the coupling efficiency, in polarization $l$, of the mode $(\omega,k)$ between bodies $m$ and $n$, $\mathbf{k}$ being the wavevector parallel to the surfaces ($k=|\mathbf{k}|$) and $\hbar \omega$ the energy carried by the mode (see Methods). In this expression  $\theta_{mn}(\omega)=\theta(\omega,T_m)-\theta(\omega,T_n)$ denotes the difference of photon mean energies at temperatures $T_m$ and $T_n$, associated to bodies $m$ and $n$, respectively. Since the gate voltages $V_{gi}$ are dynamically modulated, the temperature variation of the active membrane (pyroelectric material covered by the electrodes and the SiO$_2$ layers) is driven by the energy balance equation $c_v\delta \,dT_2/dt=\mathcal{P}_{r}(V_{gi};T_2,t)$,
where $c_v$ is the volumetric heat capacity of the membrane and $\delta$ its thickness (see Supplementary Section\,1). 

To demonstrate the potential of these converters, we first consider barium titanate (BaTiO$_3$) ceramic layer of  thickness $\delta_p=3\,\mu$m in the ferroelectric phase as the pyroelectric material\cite{Lang}
and reservoir temperatures $T_1=400\,$K and $T_3=300\,$K .  The cyclic modulation of the energy flux (Fig.\,1c) received by the active membrane induces a time variation of its temperature as shown in Fig.\,1d.
This leads to a change in the spontaneous electric dipolar moment of pyroelectric material and modifies the density of interfacial charges on the electrodes. This variation  is characterized by the pyroelectric coefficient $p=\frac{\partial P}{\partial T}$ in the direction of the poling field, $P$ being the electric polarization. If the electrodes of the pyroelectric capacitor are connected to an electric circuit, the generated current density is given by\cite{Bowen} $i_p=p(T_2)\, dT_2/dt$, which is plotted in Fig.\,1e for $f=0.2\,$kHz and $V_{g1}=V_{g2}=1\,$V (temperature-dependent specific heat, pyroelectric coefficient and permittivity of BaTiO$_3$ are given in Supplementary Section\,2). 
It is noteworthy that the thermal response $T_2(t)$ of the BaTiO$_3$ layer becomes periodic at the same frequency as the bias voltages after a transient delay (corresponding to the loss of memory of initial conditions).  
Implementing the so-called synchronized electric charge extraction (SECE) cycle\cite{Sebald2}, which consists in extracting the electric charge stored in the active material when its temperature $T_2(t)$ reaches the maximum and minimum values $T_\mathrm{max}$ and $T_\mathrm{min}$, respectively, the energy per unit surface generated during a cycle reads
\begin{equation}
W_p=\delta_p[T_\mathrm{max}p(T_\mathrm{max})-T_\mathrm{min}p(T_\mathrm{min})]\int_{T_\mathrm{min}}^{T_\mathrm{max}}\frac{p(T_2)}{\varepsilon_{33}(T_2)}dT_2,
\label{Eq:power_gen}
\end{equation}
where $\varepsilon_{33}$ denotes the permittivity of the pyroelectric material in the polarization direction (see Supplementary Section\,3 for details).
Besides, the energy dissipated to charge the graphene capacitors during a cycle is given by $W_g=\frac{1}{2}C_g(V_{g1}^2 + V_{g2}^2)$, so that the net power per unit surface delivered by a converter with operating frequency $f$ can be estimated as $\mathcal{P}=f(W_p-W_g)$. Thus, the conversion efficiency is $\eta=(W_p-W_g)/W_\mathrm{in}$, where the input energy $W_\mathrm{in}$ corresponds to the energy absorbed by the active layer during the heating stage of cycle, $W_\mathrm{in}=\delta\int_{T_\mathrm{min}}^{T_\mathrm{max}} c_v(T_2)\,dT_2$.

The ratio $W_p/W_g$ plotted in Fig.\,2a, shows that a small amount of energy is used to charge the GFETs capacitor up to a modulation frequency of about 1.5\,kHz.  This ratio is approximately constant for small frequencies, the variation amplitude of the active layer temperature having reached its upper value. Beyond this plateau, $W_p/W_g$ decreases while the variation amplitude of the temperature decreases as well. Since $W_p/W_g>1$ at kHz frequencies, the energy generated per cycle is larger than the energy used to tune the state of GFETs, demonstrating so that these devices can be self-powered throughout the conversion process. The corresponding useful (net) power $\mathcal{P}$ is shown in Fig.\,2b and the conversion efficiency in Fig.\,2c, the latter being rescaled with the Carnot efficiency $\eta_C=1-T_3/T_1$. The delivered power with this material reaches values around $10\,$mW$.$cm$^{-2}$ at frequencies of fraction of kHz.
To assess the potential of this technology with other pyroelectric materials we consider a simplified form for the generated energy\cite{Sebald2} $W_p=\delta_p (T_\mathrm{max}-T_\mathrm{min})^2p^2/\varepsilon_{33}$ when the materials properties are assumed to be independent of temperature. This expression depends on the figure of merit (FOM) $p^2/\varepsilon_{33}$ characterizing the pyroelectric performance of the material\cite{Bowen}. Parametrizing material properties with this FOM, in Figs.\,2d-f we estimate the performance of the converter for different source temperatures. The specific heat of different pyroelectric materials is set to $2.5\times10^6\,$J.\,m$^{-3}$.\,K$^{-1}$, a value which is representative of most ferroelectric materials\cite{Sebald2} as indicated in Fig.\,2e for some of them for guidance reasons. We observe that a power of a few mW$.$cm$^{-2}$ can be obtained even for small temperature differences. Moreover, neglecting the temperature dependence in $W_p$ and writing $W_g$ in terms of the carrier densities $n_{gi}$ in the graphene sheets, we find that
\begin{equation}
\frac{W_p}{W_g}= \frac{2p^2(T_\mathrm{max}-T_\mathrm{min})^2}{e^2(n_{g1}^2 + n_{g2}^2)} \frac{\delta_p\varepsilon_g}{\delta_g\varepsilon_{33}},
\end{equation}
showing a certain flexibility for those systems to achieve $W_p/W_g>1$ by an appropriate choice of thicknesses (capacitances) in the GFETs and the active layer.


To give an insight into the coupling mechanism of thermal photons in the converter, we show in Figs.\,3a-d the transmission coefficients for TM polarization (strongly dominant) of the energy carried by the electromagnetic modes $(\omega,k)$ between the active zone and the GFETs. We also show the Planck windows where the heat transfer takes place (dashed blue and red lines). For $V_{gi}=0$, we see in Figs.\,3a and 3d that the heat transfer is mainly mediated by hybridized (symmetric and antisymmetric) surface phonon polaritons (SPPs) supported by the SiO$_2$ layers on both the GFETs and the active zone, at frequencies about $0.9\times10^{14}\,$rad$.$s$^{-1}$ and $2.2\times10^{14}\,$rad$.$s$^{-1}$. When the bias voltage is switched on, the coupling of SPPs through the gap is reduced by the presence of delocalized graphene plasmon on the GFET, whose dispersion relation is shown in dashed green lines in Figs.\,3b and 3c. Consequently, the number of modes participating in the heat transfer around the SPP resonances decreases significantly when the GFET is charged. Hence, by tuning the gate voltage in the GFETs we can actively and locally (Fig.\,3e) control the near-field heat exchanges in the converter during the heating and cooling stages of the cycle (Fig.\,3f).

Implementing the SECE cycle as previously discussed is efficient when the temperature changes of ferroelectric materials take place close to their Curie temperature. Unfortunatly, when we move away from this critical temperature their pyroelectric coefficient drops sharply limiting so the electric current generation\cite{Sebald2}.
However, more efficient thermodynamic cycles can be used to improve the performances of conversion process, such as Ericsson cycles consisting in two isothermal and two isoelectric stages\cite{Olsen}. These cycles require the action of an external electric field $E$ on the pyroelectric material. By neglecting the temperature dependence of the pyroelectric coefficient in the considered working temperature range and assuming that the heat capacity does not depend on the electric field, the energy per unit surface generated with this cycle can be written as\cite{Sebald}
\begin{equation}
W_p=\delta_p (T_\mathrm{max} - T_\mathrm{min})\int_0^{E_\mathrm{max}} p(E)\,dE,
\label{Eq:power_Ericson}
\end{equation}
where $E_\mathrm{max}$ is the maximum value of the applied field (see Supplementary Section\,4 for details). 
In this case the net power  $\mathcal{P}$ and the conversion efficiency $\eta$ take the same form as previously but the input energy becomes $W_\mathrm{in}=\delta\int_{T_\mathrm{min}}^{T_\mathrm{max}} c_v(T_2)\,dT_2 + \delta_pT_\mathrm{max}\int_0^{E_\mathrm{max}} p(E)\,dE$. Indeed in Ericsson cylces  it includes also heat absorption due to electrocaloric effect at the high temperature isotherm. Hence, high performances can be achieved with materials showing a large electrocaloric activity\cite{Sebald}, as reported, for instance, in the thin film relaxor ferroelectric\cite{Mischenko} \mbox{0.90Pb(Mg$_{1/3}$Nb$_{2/3}$)O$_3$-0.10PbTiO$_3$} (also denoted as 0.9PMN-0.1PT) for a field $E_\mathrm{max}=895\,$kV$/$cm. An energy density of pyroelectric conversion $W_p/\delta_p=0.432\,$J$.$cm$^{-3}$ has been estimated\cite{Sebald} for this material, with working temperatures corresponding to $T_\mathrm{min}=338\,$K and $T_\mathrm{max}=348\,$K. This allows us to evaluate the performance of our converter under these conditions, as also considered by\cite{Fang} Fang et al.. To enhance the generated current we structure the pyroelectric material (0.9PMN-0.1PT) in a series of ten parallel thin films of thickness $300\,$nm separated by Au electrodes of thickness $50\,$nm, so that $\delta_p=3\,\mu$m, and the specific heat of 0.9PMN-0.1PT is taken as\cite{Sebald} $3\times10^6\,$J.\,m$^{-3}$.\,K$^{-1}$.
For a source and a sink at temperatures $T_1=383\,$K and $T_3=283\,$K, respectively, we show in Fig.\,4a the evolution of $T_\mathrm{max}$ and $T_\mathrm{min}$ as a function of the driving frequency $f$ in the GFETs with actuated voltages $V_{g1}=1.9\,$V and $V_{g2}=4\,$V (as shown in Fig.\,3). The proper range of working temperatures is achieved at $f=1.02\,$kHz for this configuration, but other possibilities exist because of the freedom to choose the control parameters of the device.
In Fig.\,4b we show a cut of the configuration space $(f,V_{g1},V_{g2})$ leading to the required temperature oscillations in the active zone. 
The corresponding power $\mathcal{P}$ reaches values of about $130\,$mW$.$cm$^{-2}$ with energy ratios $W_p/W_g\gg1$, which are both plotted in Figs.\,4c and 4d as a function of the frequency $f$ and actuated voltage $V_{g2}$, respectively. Moreover, ignoring the loses $W_g$, the efficiency ratio $\eta/\eta_C$ for these configurations takes a value of $3.2\,$\%. It is worthwhile to note that the power density is about 200 times larger than the result reported by\cite{Fang} Fang et al.

In conclusion, we have introduced an innovative solution to harvest energy from low-grade heat sources using pyroelectric systems driven at kHz frequencies by GFETs. Generated power densities up to 130\,mW\,cm$^{-2}$ have been predicted with relaxor ferroelectrics used in Ericsson cycles with temperatures differences of $100\,$K between the primary source and the cold sink. In addition, we have shown that the power generated by these autonomous systems surpasses the current production of NTPV devices\cite{Fiorino,Bhatt} by several orders of magnitude. Beyond its potential for near-field energy conversion, nanoscale solid-state cooling and nanoscale thermal management could also benefit from this technology.




\begin{methods}

\subsection{Energy transmission coefficients.}
In a three-body system as sketched in Fig.\,1a, the radiative heat exchange takes place, in general, through all bodies in the system, including a direct exchange between the source and the sink when the active zone is partially transparent to electromagnetic radiation. However, under the assumption that the electrodes in the active zone are opaque, meaning that these layers are optically thick, there is no direct heat exchange between the source and the sink. This amounts to consider the electrodes as semi-infinite slabs, for which the many-body energy transmission coefficients\cite{Latella} in this case reduce to
\begin{equation}
\mathcal{T}^{mn}_l =\Pi^\mathrm{pw}\frac{(1- |\rho^m_l |^2)(1- |\rho^n_l |^2) }{ \left|1- \rho^m_l\rho^n_l e^{-2i k_z d_{mn}}\right|^2}
+\Pi^\mathrm{ew}\frac{4 \mathrm{Im}( \rho^m_l ) \mathrm{Im}( \rho^n_l ) e^{- 2 \mathrm{Im}(k_z) d_{mn}}}{ \left|1- \rho^m_l \rho^n_l e^{-2 \mathrm{Im}(k_z) d_{mn}}\right|^2} 
\label{transmission_coefficients}
\end{equation}
with the source-sink coupling $\mathcal{T}^{13}_l=0$, where $d_{mn}$ is the separation distance between bodies $m$ and $n$, $k_z = \sqrt{\omega^2/c^2 -k^2}$ is the normal component of wavevector in vacuum, $\rho^m_l$ is the Fresnel reflection coefficients of body $m$, and $\Pi^\mathrm{pw}=h(\omega-ck)$ and $\Pi^\mathrm{ew}=h(ck-\omega)$ are the propagating and evanescent wave projectors, respectively, $c$ being the speed of light and $h(x)$ the Heaviside step function. Here the coefficients $\rho_l^1$ and $\rho_l^3$ correspond to the reflection coefficient of a bilayer with $n$-doped Si as substrate and a superficial layer of SiO$_2$ covered by the graphene sheet, while the coefficient $\rho_l^2$ correspond to a bilayer with Au as substrate and a superficial layer of SiO$_2$.

\subsection{Optical properties.}

For the materials in the source and sink, we consider an $n$-type heavily doped Si substrate whose dielectric properties are given by the Drude model 
\begin{equation}
\varepsilon (\omega) =\varepsilon_{\infty}-\frac{\omega_{p}^2}{\omega(\omega+i\gamma)},
\label{Drude}
\end{equation}
where\cite{Basu,Zhang} the high-frequency dielectric constant is $\varepsilon_{\infty}=11.7$, the plasma frequency is given by $\omega_p=\sqrt{Ne^2/(m^*\varepsilon_0)}$ and the scattering rate is obtained as $\gamma=e/(\mu_e m^*)$, where $\varepsilon_0$ is the vacuum permittivity, $N$ is the carrier concentration, $m^*= 0.27m_0$ is the carrier effective mass and $\mu_e$ is the carrier mobility, $m_0$ being the free electron mass. In the present study we take the carrier concentration as $N=10^{20}\,$cm$^{-3}$.
Furthermore, the dielectric permittivity of the Au electrodes in the active zone are also described by the Drude model (\ref{Drude}) with $\varepsilon_{\infty}=1$, $\gamma=5.32\times 10^{13}\,$s$^{-1}$ and $\omega_p=1.37\times 10^{16}\,$rad.\,s$^{-1}$. The dielectric permittivity of SiO$_2$ is tabulated in ref.\cite{Palik}.

The response of the graphene sheets is described in terms of a 2D frequency-dependent conductivity $\sigma(\omega)=\sigma_D(\omega)+\sigma_I(\omega)$ with intraband and interband contributions respectively given by\cite{FalkovskyJPhysConfSer08}
\begin{equation}
\begin{split}
\sigma_D(\omega)&=\frac{i}{\omega+\frac{i}{\tau}}\frac{2e^2k_BT}{\pi\hbar^2}\log\Bigl(2\cosh\frac{\mu}{2k_BT}\Bigr),\\
\sigma_I(\omega)&=\frac{e^2}{4\hbar}\Bigl[G\Bigl(\frac{\hbar\omega}{2}\Bigr)+i\frac{4\hbar\omega}{\pi}\int_0^{\infty}\frac{G(\xi)-G\bigl(\frac{\hbar\omega}{2}\bigr)}{(\hbar\omega)^2-4\xi^2}\,d\xi\Bigr],  
\end{split}
\end{equation}
where $G(x)=\sinh(x/k_BT)/[\cosh(\mu/k_BT)+\cosh(x/k_BT)]$. As these expressions show, the conductivity depends explicitly on the temperature $T$ of the graphene sheet,  its  chemical potential $\mu$ and the relaxation time $\tau$ for which we have used the value\cite{JablanPRB09} $\tau=10^{-13}\,$s. Since the graphene sheet lays on the surface of a medium with permittivity $\varepsilon(\omega)$ (this medium here is SiO$_2$), the conductivity $\sigma(\omega)$ modifies the vaccum-medium Fresnel reflection and transmission coefficients of the interface $r_l$ and $t_l$, respectively, which for the two polarizations take the form\cite{FalkovskyJPhysConfSer08,Geim3}
\begin{equation}\label{Fresnel}
\begin{split}
r_\text{TE}&=\frac{k_{z}-k_{zm}-\mu_0\sigma(\omega)\omega}{k_{z}+k_{zm}+\mu_0\sigma(\omega)\omega},\qquad
r_\text{TM}=\frac{\varepsilon(\omega) k_{z}- k_{zm}+\frac{\sigma(\omega)k_{z}k_{zm}}{\varepsilon_0\omega}}{\varepsilon(\omega) k_{z}+ k_{zm}+\frac{\sigma(\omega)k_{z}k_{zm}}{\varepsilon_0\omega}},\\
t_\text{TE}&=\frac{2 k_{z}}{k_{z}+k_{zm}+\mu_0\sigma(\omega)\omega},\qquad
t_\text{TM}=\frac{2\sqrt{\varepsilon(\omega)k_{z} }}{\varepsilon(\omega) k_{z}+ k_{zm}+\frac{\sigma(\omega)k_{z}k_{zm}}{\varepsilon_0\omega}},
\end{split}
\end{equation}
where $\mu_0$ is the vacuum permeability and $k_{zm}=\sqrt{\varepsilon(\omega) \omega^2/c^2-k^2}$ is the normal component of the wave vector in the medium. The dispersion relation of graphene plasmon is given by the zeros of the denominator of Fresnel coefficients in $TM$ polarization.
\end{methods}



\begin{addendum}
 \item This project has received funding from the European Union’s Horizon 2020 research and innovation
programme under the Marie Sklodowska-Curie grant agreement No~892718~(I.L.).
 \item[Authors contribution]
I.L. and P.B.-A. contributed equally to this work.
 \item[Competing Interests] The authors declare that they have no competing financial interests.
 \item[Correspondence] Correspondence and requests for materials should be addressed to I.L or \mbox{P.B.-A.} (email: ilatella@ub.edu; pba@institutoptique.fr).
\end{addendum}


\clearpage

\begin{figure}
\caption{\textbf{Graphene-based pyroelectric converter.} \textbf{a} Schematic illustration of the device: a pyroelectric membrane (active zone) is suspended between two GFETs (thermal reservoirs) held at two different temperatures $T_1$ (primary source) and $T_3<T_1$ (thermal sink). A modulation of bias voltages $V_{g1}$ and $V_{g2}$ applied to the GFET gates allows to oscillate the temperature $T_2$ of the membrane which generates useful power due to pyroelectric effect. The electrodes in the active zone serve to extract electric charge and apply electric field. \textbf{b} Crystallographic structure of BaTiO$_3$ showing that a permanent polarization $\textbf{P}$ exists in the $c$-direction when $T$ is smaller than the Curie temperaure $T_C$. \textbf{c} Heat flux received by the membrane during the heating-cooling steps induced by the bias voltages cycling for a layer of BaTiO$_3$ of thickness $\delta_p=3\,\mu$m. \textbf{d} Temporal evolution of the membrane while the bias voltages are modulated at frequency $f=0.2\,$kHz with turn-on values $V_{g1}=V_{g2}=1\,$V. \textbf{e} Short-circuit pyroelectric current density.}
\label{fig1}
\end{figure}

\begin{figure}
\caption{\textbf{Performances of the converter with SECE cycles.} \textbf{a} Ratio of the energy $W_p$ generated per cycle to the energy $W_g$ required to tune the gate voltages. \textbf{b} Power generated by the converter in response to the periodically varying (rectangular) bias voltage applied on the gate of the GFETs at the frequency $f$ with a primary source at $T_1=400\,$K. \textbf{c} Conversion efficiency  $\eta=(W_p-W_{g})/W_\mathrm{in}$ normalized by the Carnot efficiency $\eta_C$. \textbf{d} Time variation of the active layer temperature. \textbf{e} Generated power and \textbf{f} conversion efficiency with respect to the FOM $p^2/\varepsilon_{33}$. In all figures, $T_3=300\,$K and $V_{g1}=V_{g2}=1\,$V when these voltages are switched on.}
\label{fig2}
\end{figure}

\begin{figure}
\caption{\textbf{Coupling tunability in near-field regime.} 
\textbf{a}-\textbf{d} Transmission coefficients $\mathcal{T}^{\text{12}}_\mathrm{TM}$ and $\mathcal{T}^{\text{23}}_\mathrm{TM}$ in the $(\omega,k)$ plane between the primary source at $T_1=383\,$K and the active layer and between the active layer and the thermal sink at $T_3=283\,$K, respectively, for $T_2=343\,$K. 
Cyan curves represent the light line $\omega=ck$. 
Dashed red and blue lines denote the difference of Planck functions $\omega^2\theta_{12}$ and $\omega^2\theta_{23}$ (Planck windows), respectively, where heat transfer takes place.
Resonances around the frequencies $0.9\times10^{14}\,$rad$/$s and $2.2\times10^{14}\,$rad$/$s correspond to the symmetric and antisymmetric SPPs supported by the SiO$_2$ layers, while
dashed green lines repesent the surface plasmon dispersion relation of graphene for chemical potentials $\mu_{g1}=0.33\,$eV in \textbf{b} and $\mu_{g2}=0.48\,$eV in \textbf{c}. The anticrossing curves in \textbf{b} and \textbf{c} show the strong coupling between the SPPs of silica and the delocalized plasmon of graphene. \textbf{e} Spectrum of the heat exchange between the source and the active zone ($\varphi_{12}$) and between the active zone and the sink ($\varphi_{23}$). \textbf{f} Radiative heat flux on the active zone for the heating and cooling processes.}
\label{fig3}
\end{figure}

\begin{figure}
\caption{\textbf{Performances of the converter with Ericsson cycles.} \textbf{a} Maximum and minimum temperatures of the active zone during the modulation as a function of the frequency $f$. The active material is the relaxor ferroelectric 0.9PMN-0.1PT with an overall thickness $\delta_p=3\mu\,$m.
\textbf{b} Temperature of the active zone and cut of the configuration space $(f,V_{g1},V_{g2})$ leading to temperature oscillations with $T_\mathrm{max}=348\,$K and $T_\mathrm{min}=338\,$K. \textbf{c} Generated power as a function of the turn-on voltage $V_{g2}$.
\textbf{d} Generated power as a function of the modulation frequency $f$. The insets in \textbf{c} and \textbf{d} show the ratio of the energy $W_p$ generated per cycle to the energy $W_g$ required to tune the gate voltages.
}
\label{fig4}
\end{figure}

\clearpage   
\begin{figure}
\includegraphics[width=\textwidth]{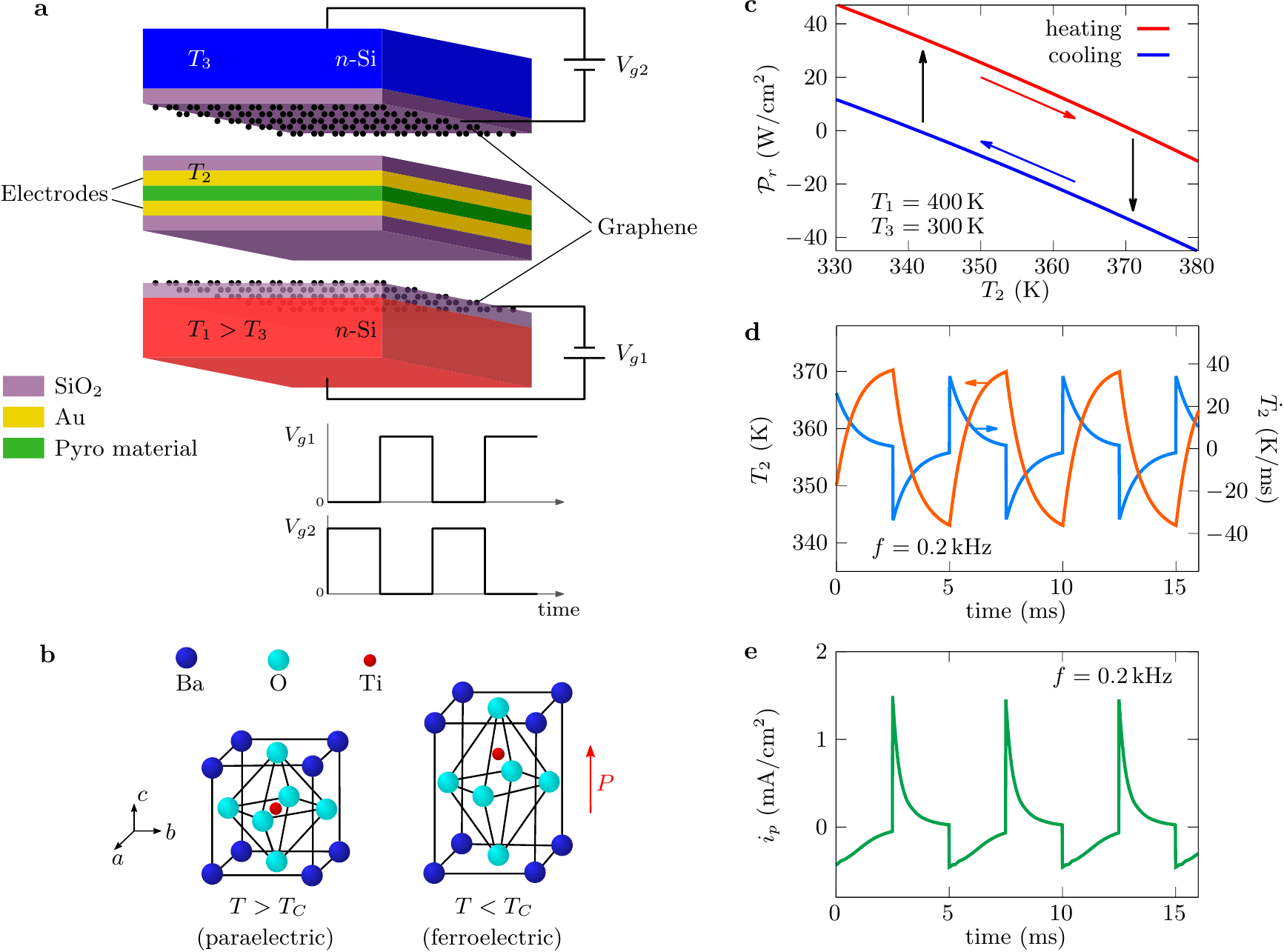}
\begin{center}
Figure 1 
\end{center}
\end{figure}
\clearpage

\begin{figure}
\includegraphics[width=\textwidth]{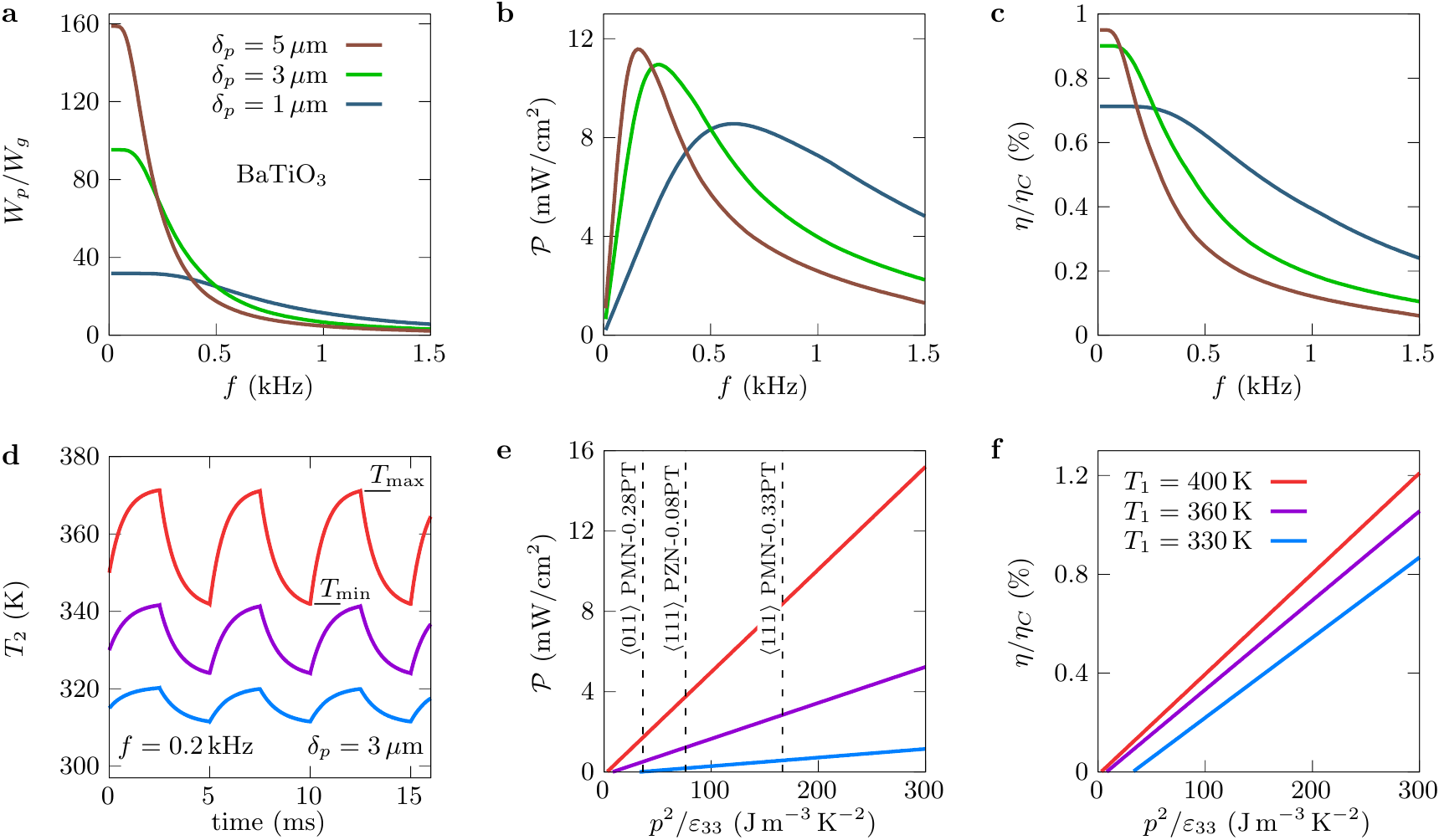}
\begin{center}
Figure 2 
\end{center}
\end{figure}
\clearpage

\begin{figure}
\includegraphics[width=\textwidth]{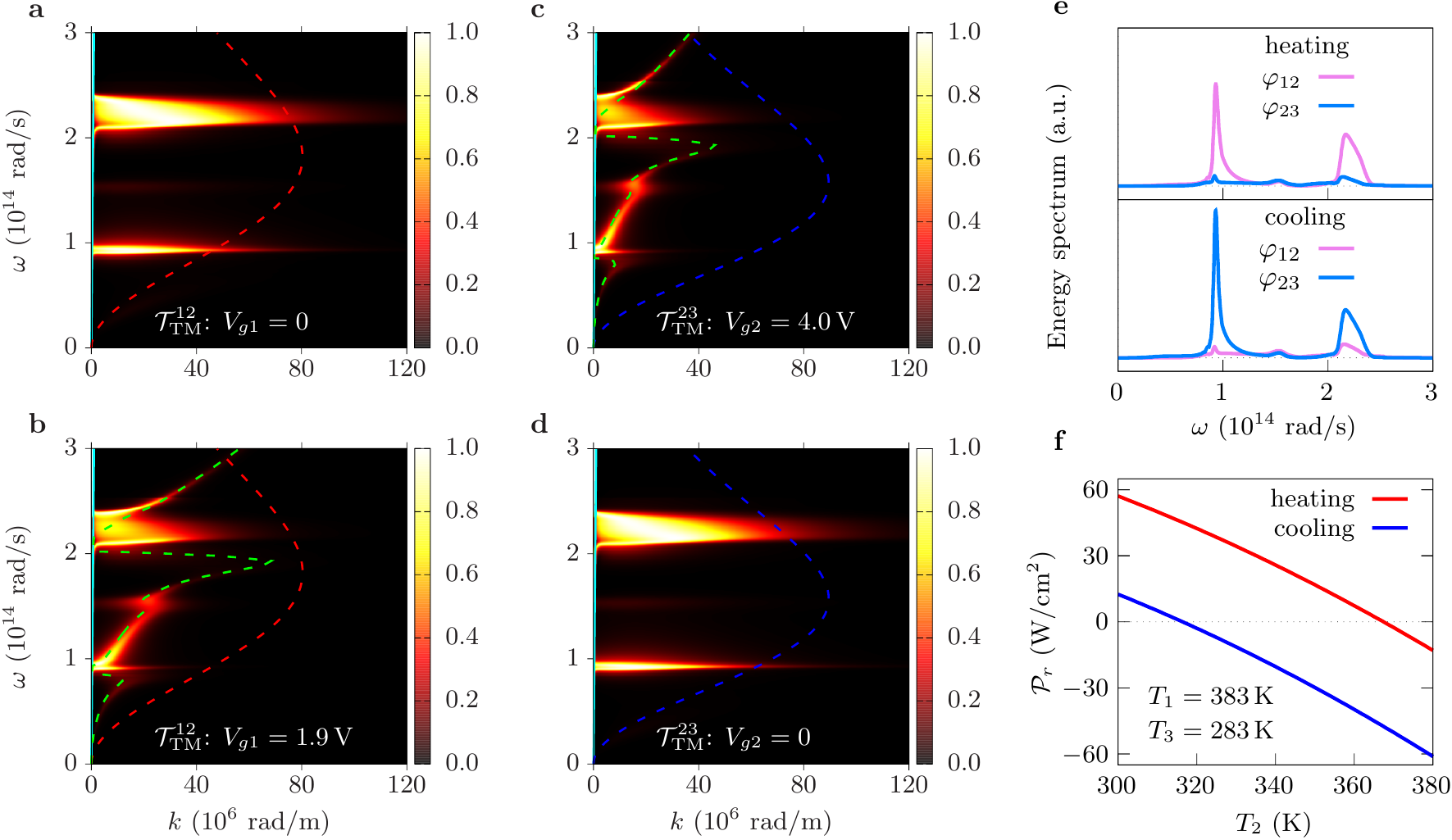}
\begin{center}
Figure 3
\end{center}
\end{figure}
\clearpage

\begin{figure}
\includegraphics[width=\textwidth]{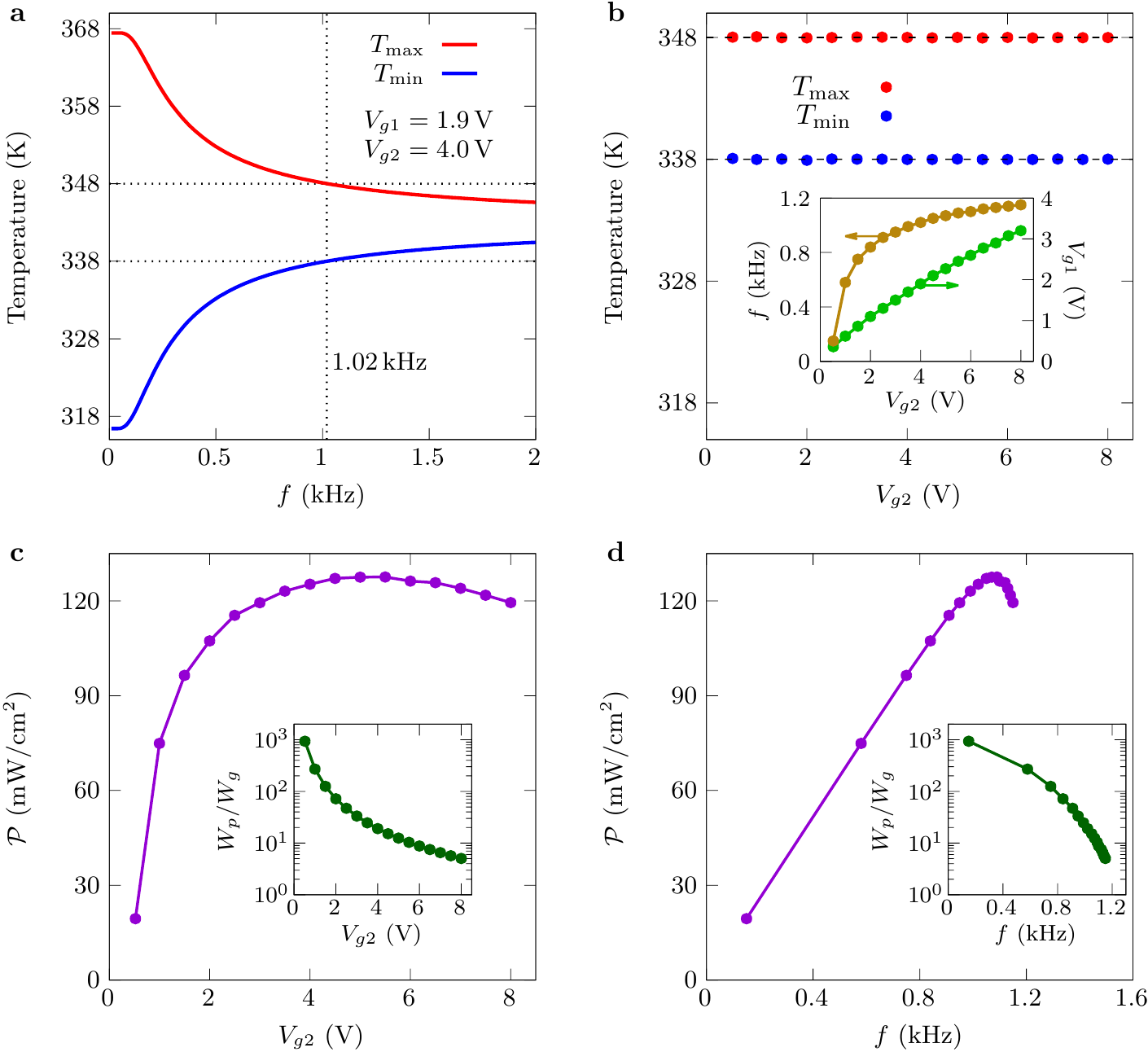}
\begin{center}
Figure 4
\end{center}
\end{figure}
\clearpage

\end{document}